%% file: cos-post_perez_astroph.tex
\documentclass[graybox]{svmult}

\usepackage{mathptmx}       % selects Times Roman as basic font
\usepackage{helvet}         % selects Helvetica as sans-serif font
\usepackage{courier}        % selects Courier as typewriter font
\usepackage{type1cm}        % activate if the above 3 fonts are
\usepackage{makeidx}         % allows index generation
\usepackage{graphicx}        % standard LaTeX graphics tool
\usepackage{multicol}        % used for the two-column index
\usepackage[bottom]{footmisc}% places footnotes at page bottom

\makeindex             % used for the subject index
\begin{document}

\title*{Search for $H_{\alpha}$ emitters in Galaxy Clusters
with Tunable Filters 
}
\author{Ricardo P\'erez Mart\'inez, Miguel S\'anchez Portal, Jordi Cepa Nogu\'e, \'Angel Bongiovanni  \and Ana P\'erez Garc\'ia.
}
%\author{Ricardo P\'erez Mart\'inez}
%\institute{Ricardo P\'erez Mart\'inez \at INSA/ESAC \email{ricardo.perez@sciops.esa.int}}
%\author{Miguel S\'anchez Portal}
%\institute{Miguel S\'anchez Portal \at INSA/ESAC }

\institute{Ricardo P\'erez Mart\'inez \at INSA/ESAC \email{ricardo.perez@sciops.esa.int},\\\\
 Miguel S\'anchez Portal \at INSA/ESAC.\\\\ Jordi Cepa Nogu\'e, \at U. de La Laguna and IAC.\\\\ \'Angel Bongiovanni \at IAC.\\\\ Ana P\'erez Garc\'ia \at IAC.  }
% \short

\authorrunning{P\'erez-Mart\'inez, R. \textit{et al.}}

\maketitle

\abstract*{The studies of the evolution of galaxies in Galaxy Clusters have as a traditional complication the difficulty in establishing cluster membership of those sources detected in the field of view. The determination of spectroscopic redshifts involves long exposure times when it is needed to reach the cluster peripherical regions of/or clusters at moderately large redshifts, while photometric redshifts often present  uncertainties too large to offer significant conclusions. The mapping of the cluster of galaxies with narrow band tunable filters makes it possible to reach large redshifts intervals with an accuracy high enough to establish the source membership of those presenting emission/absorption lines easily identifiable, as $H_{\alpha}$. Moreover, the wavelength scan can include other lines as [NII], [OIII] or $H_{\beta}$ allowing to distinguish those sources with strong stellar formation activity and those with an active galactic nuclei. All this makes it possible to estimate the stellar formation rate of the galaxies observed. This, together with ancillary data in other wavelengths may lead to a good estimation of the stellar formation histories. It will shed new light over the galaxy evolution in clusters and will improve our understanding of galaxy evolution, especially in the outer cluster regions, usually less studied and with significant unexploited data that can not be correctly interpreted 
without redshift determination.
}

\abstract{The studies of the evolution of galaxies in Galaxy Clusters have as a traditional complication the difficulty in establishing cluster membership of those sources detected in the field of view. The determination of spectroscopic redshifts involves long exposure times when it is needed to reach the cluster peripherical regions of/or clusters at moderately large redshifts, while photometric redshifts often present  uncertainties too large to offer significant conclusions. The mapping of the cluster of galaxies with narrow band tunable filters makes it possible to reach large z intervals with an accuracy high enough to establish the source membership of those presenting emission/absorption lines easily identifiable, as $H_{\alpha}$. Moreover, the wavelength scan can include other lines as NII, OIII or $H_{\beta}$ allowing to distinguish those sources with strong stellar formation activity and those with an active galactic nuclei. All this makes it possible to estimate the stellar formation rate of the galaxies observed. This, together with ancillary data in other wavelengths may lead to a good estimation of the stellar formation histories. It will shed new light over the galaxy evolution in clusters and will improve our understanding of galaxy evolution, especially in the outer cluster regions, usually less studied and with significant unexploited data that can not be correctly interpreted 
without redshift determination.
}

\section{Introduction}
\label{sec:1}
The influence of the environment on galaxy evolution has been long discussed and has been generally agreed that cluster and field galaxies evolve differently. Various effects have been addressed to explain this, though the specific processes involved are far from being understood.

An outstanding parameter to characterize the galaxy evolution is the stellar formation rate (SFR) and history (SFH) of the object as well as the eventual existence of an AGN. Various SFR indicators exist at different spectral ranges, with significant discrepancies among the results obtained from different ones. Though there are several empirical and theoretical laws between optical (e.g. $H_{\alpha}$) and MIR (e.g 15 microns from ISO observations, (Elbaz \textit{et al}. 2002 [1] ) new ones are needed to take into account the rather large number of Spitzer sources detected at new available wavelengths. Moreover, the fact that the number of MIR cluster sources that also exhibits X-Ray emission is not significant (Martini \textit{et al.} 2006 [2]) indicates either a lack of AGNs in this cluster or their heavy absortion at the observed wavelength possibly due to galaxy environment. Diagnosis based in $H_{\alpha}$ and [NII] will make it possible to distinguish between stellar formation activity or AGN powered MIR emission.(Hopkins \textit{et al.} 2003 [3]), as well as helping to detect AGNs whose X-ray emission is confused in the X-ray luminous intracluster medium.

\section{The Galaxy Cluster Abell 2219}
\label{sec:11}
A2219 is a massive galaxy cluster at z=0.2256 with a virial radius of 3 Mpc. The mass distribution traced by X-ray observation shows a SE-NW elongation consistent with an advance-phase merging process (Boschin \textit{et al}, 2004 [4]). We have also observed this elongation in Spitzer MIPS and IRAC pointings. 

\section{The idea behind}
\label{sec:2}
By performing 
 a deep H$\alpha$ and [NII] survey in the galaxy cluster Abell 2219, we meant 
 to establish new relations between the SFRs that we have obtained at 24 $\mu$m using 
 SPITZER/MIPS observations and those deduced at optical wavelengths 
(H$\alpha$). The current broadly accepted relations are based on a significantly 
lower number of MIR emitters detected by ISO at shorter wavelengths. They would also make it possible to distinguish 
between stellar formation and AGN powered MIR emission, by means of standard [NII]$\lambda$ 6583/H$\alpha$ ratio 
diagnostics, helping to map the AGN distribution within the galaxy cluster. It is worth to notice that this 
procedure can yield the detection of  new AGNs in the cluster, since the hot intracluster medium will 
not dilute the H$\alpha$ and [NII] emission of individual sources as it does in the X-ray range. 
Besides the already known cluster members, the existence of the H$\alpha$+[NII] feature would  allow us to 
establish the cluster membership of objects with unknown or uncertain redshift, therefore constraining their 
 infrared luminosities and SFRs obtained with these and previously analaysed optical and MIR data. 

\section{The Observations}
\label{sec:3}
% Always give a unique label
% and use \ref{<label>} for cross-references
% and \cite{<label>} for bibliographic references
% use \sectionmark{}
% to alter or adjust the section heading in the running head

We used CAFOS, at the Calar-Alto Astronomical Observatory to obtain deep H$\alpha$ and [NII] flux maps on Abell 2219. CAFOS ( = Calar Alto Faint Object Spectrograph ) is a focal reducer which changes the 2.2m telescope's f-ratio from f/8 to f/4.4. 
In its standard configuration CAFOS is equipped with a 2048 × 2048 pixel blue sensitive CCD, either with a SITe chip with 24 mm pixels or a LORAL chip with 15 $\mu$m pixels. This last one was used during the experiment, providing a field of view of 11x11 $arcsec^2.$
The instrument was configured with a Fabry-Perot etalon. This enabled narrow band images in the 600 - 1000 nm region at a resolution of R = $\lambda$/D$\lambda$ = 500.

Tuning of the etalon enables imaging spectroscopy in the entire field, leaving to the observer the decision of what spectral region to scan, optimizing the exposure times and avoiding the confusion of sources typical of integral field spectroscopy in this kind of high density regions. This is particularly interesting when searching for a specific spectral line (or set of lines) at an unknown redshift.
\begin{figure}[!h]
%\sidecaption
% For example, with the graphicx style use
%\includegraphics[width=0.4\linewidth,angle=0,clip=true]{A2219raw804p4.ps}
\includegraphics[width=0.5\linewidth, angle=0,clip=true]{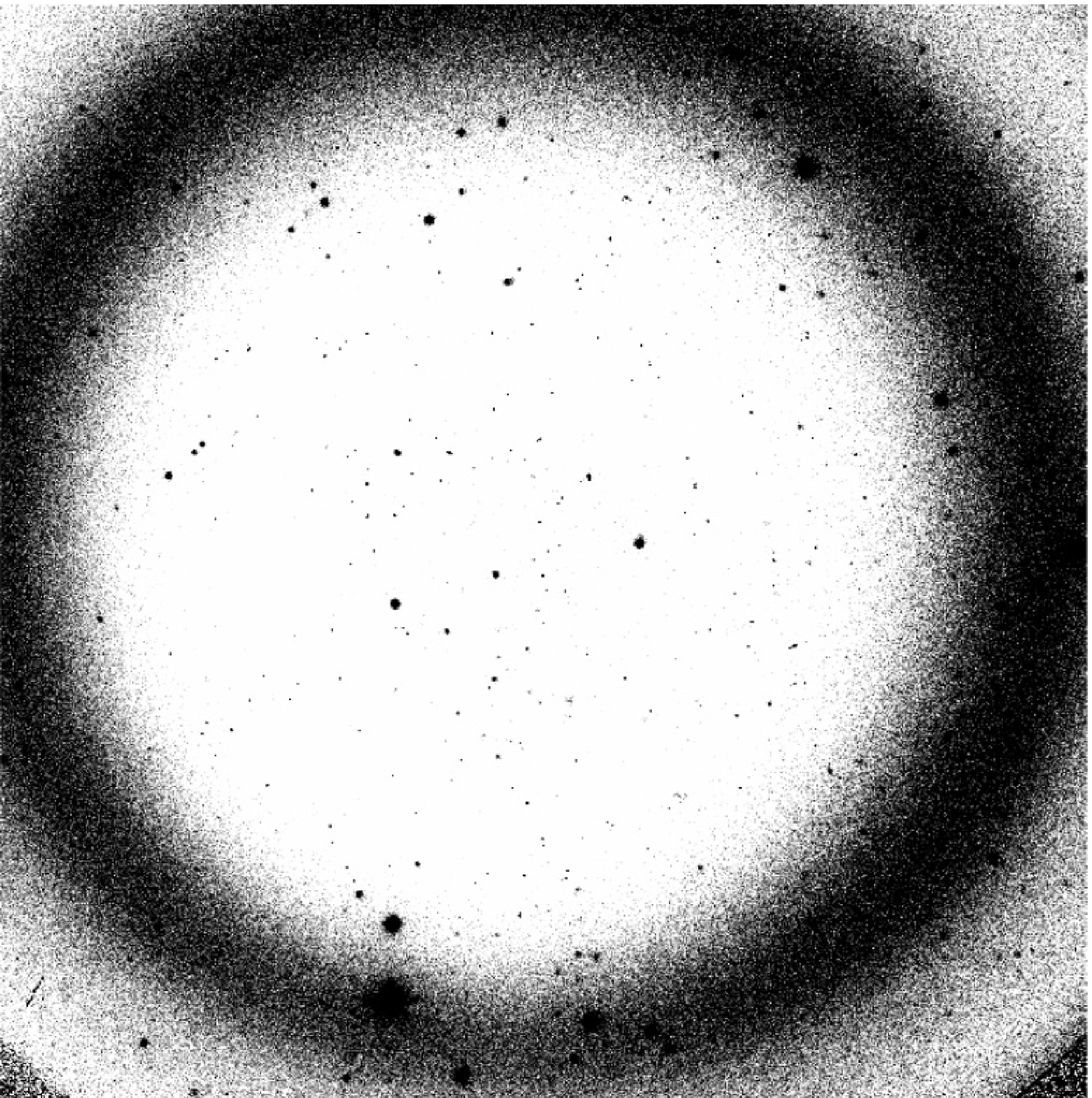}
\hspace{0.5 cm}
\includegraphics[width=0.5\linewidth, angle=0,clip=true]{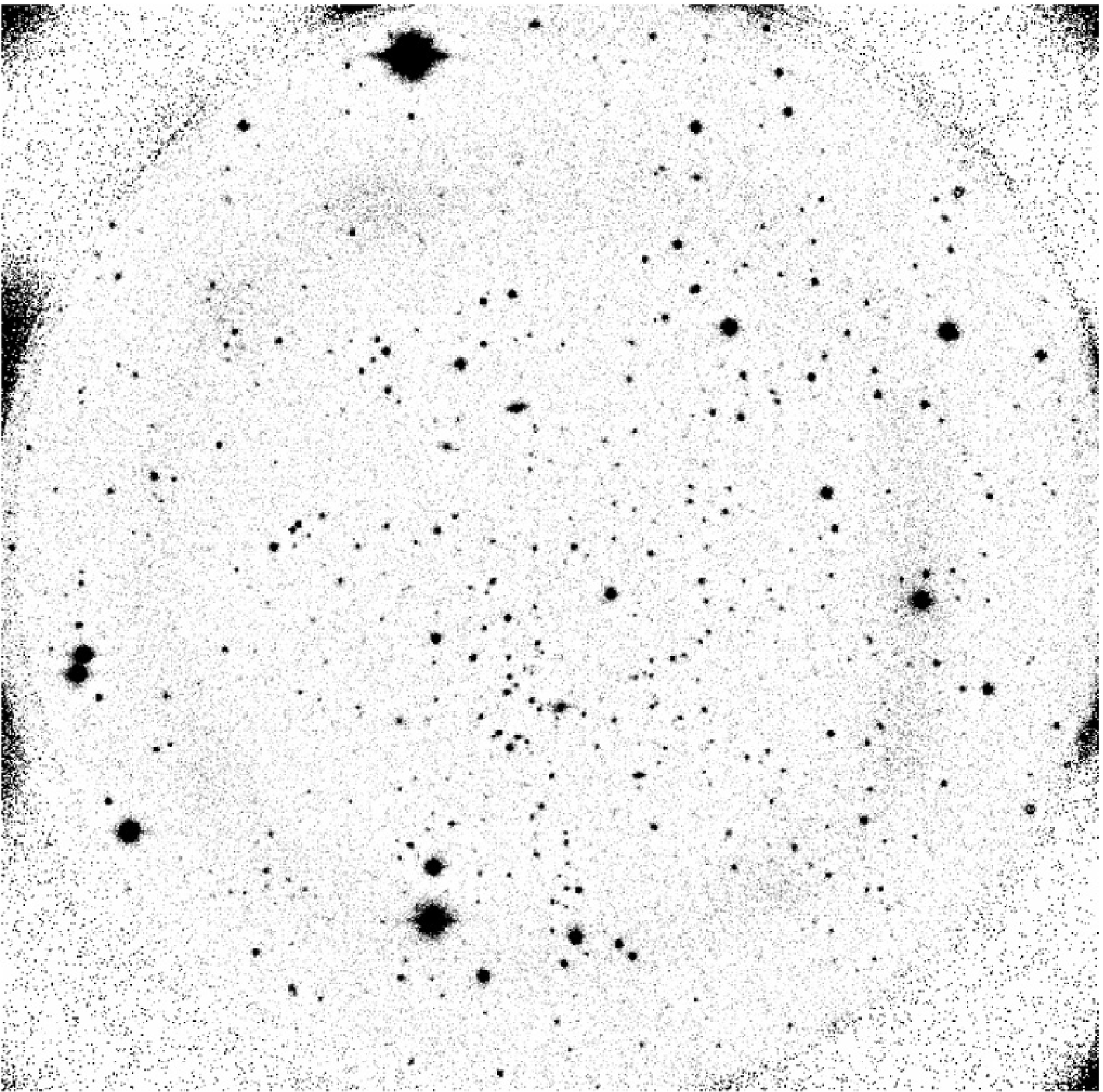}

\caption{
\textit{Left:} 1000\,sec raw image of the H$\alpha$ line at the nominal cluster redshift
($\lambda$\,=\,8044\AA)  of A2219 obtained during the Spring '07 campaign. The ring--shaped feature
is created by a sky emission line.
\textit{Right:} Composite image mapping the H$\alpha$ line at the nominal cluster redshift 
($\lambda$\,=\,8044\AA)  obtained after processing the raw images with the \texttt{TFRED} IRAF package for Fabry--Perot
image reduction. This package has been written by D.H.Jones (AAO) and modified by our team. A
stack of 11\,$\times$\,1000\,sec frames were aligned and added up. Notice that the sky line has been almost completely removed. A large number of 
H$\alpha$ emitters candidates are observed. It is needed to complete the planned program  in order to fully  
confirm their nature and cluster membership.
}

\end{figure}

\subsection{Data reduction}
One of most important tasks with dealing with tunable filter data is the removal of the sky lines, which appear as rings in the raw frames.
A complete set of tools were developed by the OSIRIS team based on a previous work of D. Heath Jones [5] and lent to us as part of
the collaboration (A. Bongiovanni \textit{priv comm}).

A good example of this is presented in \textit{figure 1, Left}, where the ring produced by a moderately weak sky feature is clearly seen. \textit{Figure 1, Right} shows the same frame after the data reduction, with the sky ring practically removed.


%\begin{figure}[htpb]
%\begin{minipage}[t]{0.5\linewidth}
%\centering
%\includegraphics[width=1\linewidth, angle=0,clip=true]{A2219raw.eps}
%\caption{1000\,sec raw image of the H$\alpha$ line at the nominal cluster redshift
%($\lambda$\,=\,8044\AA)  of A2219 obtained during the Spring '07 campaign. The ring--shaped feature
%is created by a sky emission line.}
%\end{minipage}
%\hspace{0.5 cm}
%\begin{minipage}[t]{0.5\linewidth}
%\centering
%\includegraphics[width=1\linewidth, angle=0,clip=true]{A2219cor.eps}
%\caption{Composite image mapping the H$\alpha$ line at the nominal cluster redshift 
%($\lambda$\,=\,8044\AA)  obtained after processing the raw images with the \texttt{TFRED} IRAF package for Fabry--Perot
%image reduction. This package has been written by D.H.Jones (AAO) and modified by our team. A
%stack of 11\,$\times$\,1000\,sec frames were aligned and added up. Notice that the sky line has been almost completely removed. A large number of 
%H$\alpha$ emitters candidates are observed. It is needed to complete the planned program  in order to fully  
%confirm their nature and cluster membership.}
%\end{minipage}
%\end{figure}

%
\section{Preliminary results and future work}

Hundreds of candidates to $H_{\alpha}$ emitters have been extracted from the various frames, each scanned at different central wavelenghts. The population obtained includes so galaxies with radial velocities spannig from the extremes to the center of the velocity dispersion distribution. The planned follow up with spectroscopic observation will give us important information about metallicities and extintion, to have a more accurate idea of the characteristics of the emitters in the cluster. Relating this to the position of the source in the velocity dispersion distribution will shed new light of how and where the transformation from field to cluster galaxy takes place.

\input{referenc}

\end{document}

%% file: referenc.tex
%%%%%%%%%%%%%%%%%%%%%%%% referenc.tex %%%%%%%%%%%%%%%%%%%%%%%%%%%%%%
% sample references
% %
% Use this file as a template for your own input.
%
%%%%%%%%%%%%%%%%%%%%%%%% Springer-Verlag %%%%%%%%%%%%%%%%%%%%%%%%%%
%
% BibTeX users please use
% \bibliographystyle{}
% \bibliography{}
%
%\biblstarthook{}